\documentclass[authoryear,preprint,review,12pt]{elsarticle}

\usepackage{graphics}

\usepackage{graphicx}

\usepackage{epsfig}

\usepackage{amssymb}

\usepackage{amsthm}

\usepackage{lineno}

\journal{New Astronomy}

\begin{document}

\begin{frontmatter}



\title{Absolute Properties of An Overcontact Binary HH\,Boo}


\author{H.A. Dal\corref{cor1}}
\ead{ali.dal@ege.edu.tr}
\cortext[cor1]{Corresponding author}

\author{E. Sipahi}

\address{Ege University, Science Faculty, Department of Astronomy and Space Sciences, 35100 Bornova, \.{I}zmir, Turkey}

\begin{abstract}

We obtained multi-colour light curves of HH\,Boo. We analysed the orbital period variation of the system. The analysis indicated that there is possible mass transfer from the second component to the primary or mass loss with $-5.04\times10^{-7}$ $M_{\odot}$ per year. Re-analysing the available radial velocity curve, we analysed the light curves. The inclination ($i$) of the system was found to be 69$^\circ$.71$\pm$0$^\circ$.16, while the semi-major axis ($a$) was computed as 2.246$\pm$0.064 $R_{\odot}$. The mass of the primary component was found to be 0.92$\pm$0.08 $M_{\odot}$, while it was obtained as 0.58$\pm$0.06 $M_{\odot}$ for the secondary component. The radius of the primary component was computed as 0.98$\pm$0.03 $R_{\odot}$, while it was computed as 0.80$\pm$0.02 $R_{\odot}$ for the secondary component. We demonstrated that HH\,Boo should be a member of the A-type subclass of W UMa binaries.

\end{abstract}

\begin{keyword}

techniques: photometric --- (stars:) binaries: eclipsing --- stars: late-type --- stars: individual: (HH Boo)

\end{keyword}

\end{frontmatter}



\section{Introduction}
HH\,Boo (GSC\,03472-00641) is classified as an eclipsing binary of W UMa type (a contact binary) in the SIMBAD Database. For the first time, the system was listed in the TYCHO-2 Catalogue by \citet{Hog00}, who gave $V=11^{m}.32$ and $(B-V)=0^{m}.45$ for the system. Its variability nature was found by \citet{Mac03}, who obtained the first light curve and gave the light elements as following: $T_{0}=2452764.50965$ and $P=0^{d}.318618$. Considering the spectra of the system, \citet{Mac03} indicated that the system should be from the spectral type G5III. Then, \citet{MaL04} obtained the radial velocity curve of the system. They found that mass ratio is 0.633$\pm$0.042, while $M_{1}sin^{3}i=0.78\pm0.08$ $M_{\odot}$ and $M_{2}sin^{3}i=0.49\pm0.05$ $M_{\odot}$. Although \citet{Mac03} have obtained a light curve, there is no light curve analysis given in the literature. However, many minima times have been obtained along the years. Most of them are listed in the modern database of O-C Gateway \citep{Pas06}, while some others are given by \citet{MaK04}, \citet{MaN04}, \citet{Bra08}, \citet{Hub09}, \citet{Hub10}, \citet{Wal10}.

According to \citet{Amm06}, who determined new temperatures and metallicities for more than 100,000 FGK dwarfs, the temperature of the system is 5699 K, while $[Fe/H]$ was found to be -0.57. The distance is 54 pc.

In the literature, there are several systems similar to HH\,Boo \citep{Ess10}. In this study, we obtained multi-colour light curves. Analysing the orbital period variation, we adjusted the light elements of the system and derived the mass transfer rate between the components. Then, re-analysing the radial velocity obtained by \citet{MaL04}, we analysed the multi-colour light curve using the parameters such as mass ratio and semi-major axis.

\section{Observations}
Observations of the system were acquired with a thermoelectrically cooled ALTA $U+47$ 1024$\times$1024 pixel CCD camera attached to a 35 cm - Schmidt - Cassegrains - type MEADE telescope at Ege University Observatory. The observations were continued in BVR bands in the observing season  2011. Some basic parameters of program stars are listed in Table 1, in which the brightness and colours were taken from the General Catalogue of Variable Stars \citep{Kho88}.

Although the program and comparison stars are very close on the sky, differential atmospheric extinction corrections were applied. The atmospheric extinction coefficients were obtained from observations of the comparison stars on each night. Heliocentric corrections were also applied to the times of the observations. The mean averages of the standard deviations are 0$^{m}$.023, 0$^{m}$.011, and 0$^{m}$.010 for observations acquired in the BVR bands, respectively. To compute the standard deviations of observations, we used the standard deviations of the reduced differential magnitudes in the sense comparisons (GSC\,03472-00043) minus check (GSC\,03472-01201) stars for each night. There was no variation observed in the standard brightness comparison stars.

\section{Analysis of Orbital Period Variation}
There are several minima times of the system in the literature \citep{Pas06, Hub10}. In the analysis, we used nor visual observations neither the minima times with large error. We also obtained new 15 minima times. All the minima times used in this study are listed in Table 2. The standard deviation of each one obtained in this study is given in the brackets near its related digits in Table 2.

Using the regression calculation, investigations demonstrated that the main variation could be described by a downward parabolic curve, which must be caused by possible mass transfer from the second component to the primary or mass loss. Therefore, the main period variation was represented by the  quadratic light elements, which are given by Equation (1).

\begin{center}
\begin{equation}
Min~I~(Hel.)~=~24~54912.3236(2)~+~0^{d}.318666(1)~\times~E-1.3349(1)\times10^{-11}~\times~E^{2}
\end{equation}
\end{center}
where the standard deviations of each coefficient and each constant are given in the brackets near their related digits.

The period variation and its quadratic fit are shown in Figure 1. In the analyses, the weighted sum of the squared residuals, $\Sigma w(O-C)^{2}$, was found to be 0.000038 $day^{2}$. Considering the quadratic term ($Q$), the parameter of the period variation ($dP/dt$) was found to be -9.60$\times10^{-7}$ $yr^{-1}$. The relation between the parameter of the period variation ($dP/dt$) and the rate of mass transfer was determined with Equation (2) by \citet{Pri85}:

\begin{center}
\begin{equation}
\dot{m}~=~\frac{1}{3}~\times~\frac{M_{1}~\times~M_{2}}{M_{1}~-~M_{2}}~\times~\frac{\dot{P}}{P}
\end{equation}
\end{center}
where $\dot{m}$ is mass transfer rate per year. Using this equation, the mass transferring from the secondary component to the primary was found to be $-5.04\times10^{-7}$ $M_{\odot}~yr^{-1}$.

\section{ Radial Velocity and Light Curve Analyses}
We took the radial velocity curve of the system from \citet{MaL04}. In order to adjust their radial velocity solution, we used the Spectroscopic Binary Solver software \citep{Joh04} and re-analysed the radial velocity curve. In the analysis, we used the orbital period adjusted by the $O-C$ analyses in this study, and we fixed it. The results are listed in Table 3. The solution generally gave the same values obtained by \citet{MaL04} with the little differences. The theoretical radial velocity curves derived by the Spectroscopic Binary Solver software are shown in Figure 2.

HH\,Boo is classified as an eclipsing binary of W UMa type in the SIMBAD Database, and it is more likely a contact binary system. In this respect, we analysed the light curves obtained in the BVR bands with using the PHOEBE V.0.31a software \citep{Prs05}, which is used in the version 2003 of the Wilson-Devinney Code \citep{Wil71, Wil90}. We tried to analyse the light curves with different modes, such as the "overcontact binary not in thermal contact", "semi-detached system with the primary component filling its Roche-Lobe", and "double contact binary" modes. The initial analyses demonstrated that an astrophysical acceptable result can be obtained if the analysis is carried out in the "overcontact binary not in thermal contact" mode, while no acceptable results in the astrophysical sense could be statistically obtained in all the others modes.

According to the radial velocity analysis, the mass ratio of the system should be 0.632$\pm$0.038. \citet{Amm06} determined the temperature of the system as 5699 K. Thus, the temperature of the primary component was fixed to 5699 K in the analyses, and the temperature of the secondary was taken as a free parameter. Considering the spectral type corresponding to this temperature, the albedos ($A_{1}$ and $A_{2}$) and the gravity darkening coefficients ($g_{1}$ and $g_{2}$) of the components were adopted for the stars with the convective envelopes \citep{Luc67, Ruc69}. The non-linear limb-darkening coefficients ($x_{1}$ and $x_{2}$) of the components were taken from \citet{Van93}. In the analyses, the fractional luminosity ($L_{1}$) of the primary component and the inclination ($i$) of the system were also taken as the adjustable free parameters.

The parameters derived from the analyses are listed in Table 4, while the synthetic light curves are shown in Figure 3. In addition, using the parameters obtained from the light curve analysis, we also derived the Roche geometry of the system that is shown in Figure 4.

Using the radial velocity curves, the mass ratio of the system ($q$) was obtained as 0.632$\pm$0.038, and the semi-major axis ($a$) was found to be 2.246$\pm$0.064 $R_{\odot}$. Considering both the radial velocity curve solution and the inclination ($i$) of the system found from the light curve analysis, the masses were found to be 0.92$\pm$0.08 $M_{\odot}$ for the primary component and 0.58$\pm$0.06 $M_{\odot}$ for the secondary component. Considering the semi-major axis, the radius of the primary component was computed as 0.98$\pm$0.03 $R_{\odot}$, while it was computed as 0.80$\pm$0.02 $R_{\odot}$ for the secondary component. In addition, the luminosity of the primary component was computed as 0.91$\pm$0.08 $L_{\odot}$, and it was computed as 0.47$\pm$0.03 $L_{\odot}$ for the secondary component.

In order to test whether the absolute parameters are generally acceptable in the astrophysical sense, or not, we compared the components with other systems in the mass-radius ($M-R$), mass-luminosity ($M-L$), and luminosity-effective temperature ($T_{eff}-L$) planes. All the comparisons are shown in Figure 5. In the figure, the lines represent the ZAMS theoretical model developed for the stars with $Z=0.02$ by \citet{Gir00}, while dashed lines represent the TAMS theoretical model. The filled circles represent the primary components, while the open circles represents the secondary ones. The components of HH\,Boo are located together with some samples of its analogues, such as YY\,CrB, DN\,Boo, CK\,Boo, $\epsilon$\,CrA, FG\,Hya, TV\,Mus, AW\,UMa, GR\,Vir, V776\,Cas.  The sample systems were taken from \citet{Ess10}, and they are shown in purple colour, while HH\,Boo is shown in black colour in Figure 5.

\section{Discussion}
In this study, we tried to determine the nature of an eclipsing binary system HH\,Boo. The analysis of orbital period variation indicates possible mass transfer from the second component to the primary and/or possible mass loss from the system. The size of transferring or loosing mass was computed as $-5.04\times10^{-7}$ $M_{\odot}$ per year, which causes a decreasing in the orbital period due to the angular momentum loss. In fact, the parameter of the period variation was found to be -9.60$\times10^{-7}$ $yr^{-1}$. We adjusted the orbital period as $0^{d}.318666$. It is well known that several binaries of W UMa type such as YY\,CrB \citep{Ess10}, BS\,Cas \citep{Yan08}, VZ\,Tri \citep{Yan10}, exhibit large the period variation due to the large mass transfer.

We also re-analysed available radial velocity curve. Our solution seems to be a little bit different according to the results found by \citet{MaL04}. The differences should be caused due to the orbital period used in this study. Here, the adjusted period is a bit different from the one used by \citet{MaL04}.

For the first time in the literature, we analysed light curves of the system. The inclination ($i$) of the system was found to be 69$^\circ$.71$\pm$0$^\circ$.16, while the temperature of the secondary component was found to be 5352$\pm$15 K from the analysis. The fractional radii were found to be $r_{1}=0.435\pm0.001$ for the primary component and $r_{2}=0.354\pm0.001$ for the secondary one. In this case, the sum of fractional radii was computed as $r_{1}+r_{2}\simeq0.80$. Thus, HH\,Boo seems to be in agreement with \citet{Kop56}'s criteria for overcontact systems. The $O-C$ analysis indicates that the orbital period as $0^{d}.318666$. In addition, the temperature of the primary component is 5699 K, while the secondary one is 5352 K. Although some W UMa type binaries have components with some different surface temperature, they generally have the same surface temperature. Here, the primary component of HH\,Boo is a little bit hotter than the secondary one. The $O-C$ analysis indicates possible mass transfer from the second component to the primary. This case should be the reason of the hotter primary component. Considering some characteristics of the system such as the short orbital period, small mass ratio, hotter primary component and ect., HH\,Boo seems to be in agreement with the members of the A-type subclass of W UMa binaries \citep{Ber05, Ruc85}.

As it is seen from Figure 5, comparing HH\,Boo with its analogues in some planes, such as $M-R$, $M-L$, and $T_{eff}-L$ planes, demonstrated that the components of the system are in agreement with their analogues. On the other hand, both components are seen closer to each other in the figures. If the result obtained from the $O-C$ analysis is taken into account, it is possible that the secondary component will continue to lose its mass, while the primary will collect. Therefore, both components will have been separated from each other in the planes shown in Figure 5. However, they will get closer to each other in their orbits as it is in the models of \citet{Roc02}.

\section*{Acknowledgments} The author acknowledges the generous observing time awarded to the Ege University Observatory. We also thank the referee for useful comments that have contributed to the improvement of the paper.

\clearpage

\begin{figure*}
\hspace{2.2 cm}
\includegraphics[width=15cm]{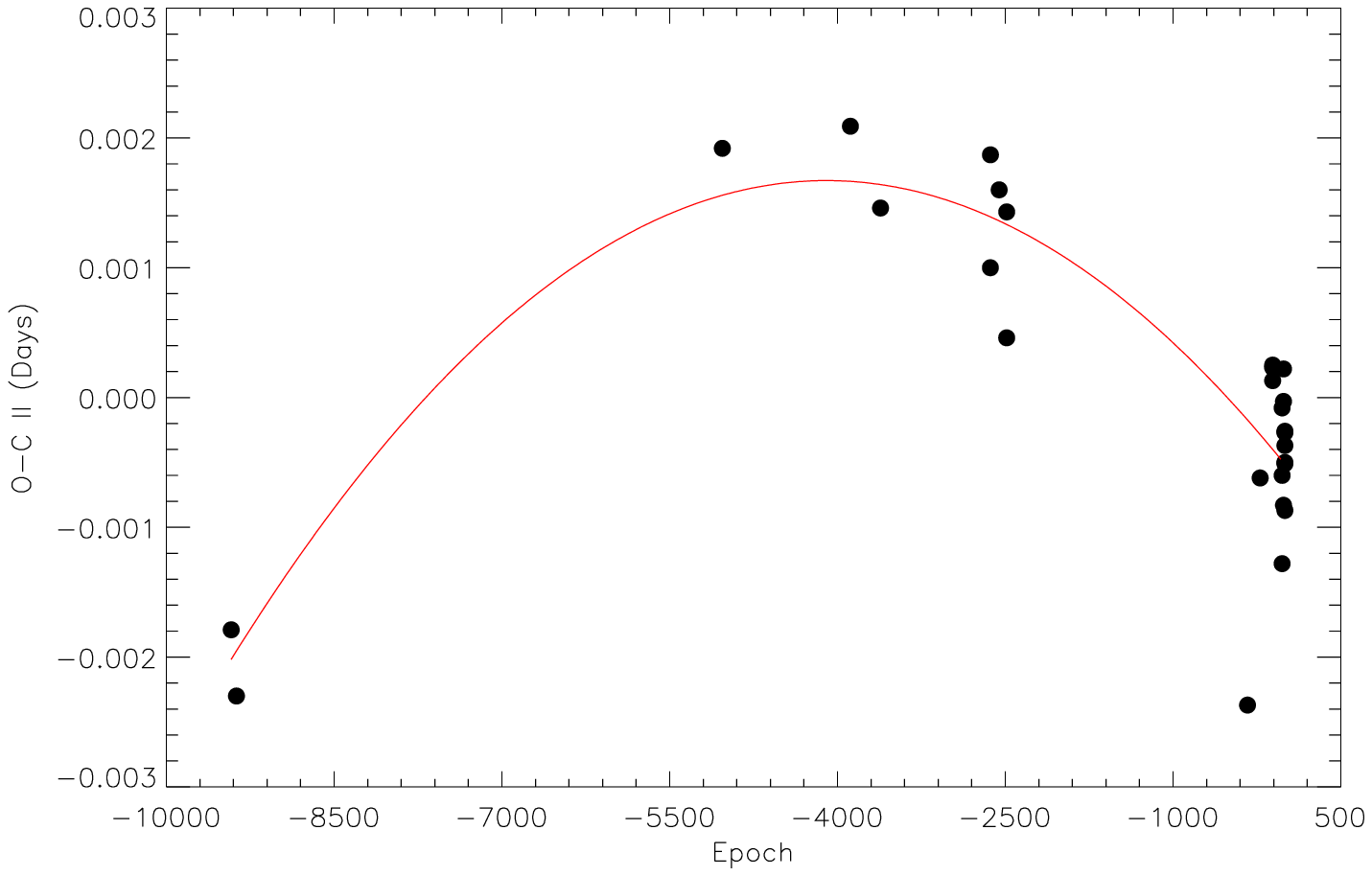}
\vspace{0.1 cm}
\caption{The O-C variation of HH\,Boo.}
\label{Fig.1}
\end{figure*}

\begin{figure*}
\hspace{1.0 cm}
\includegraphics[width=16cm]{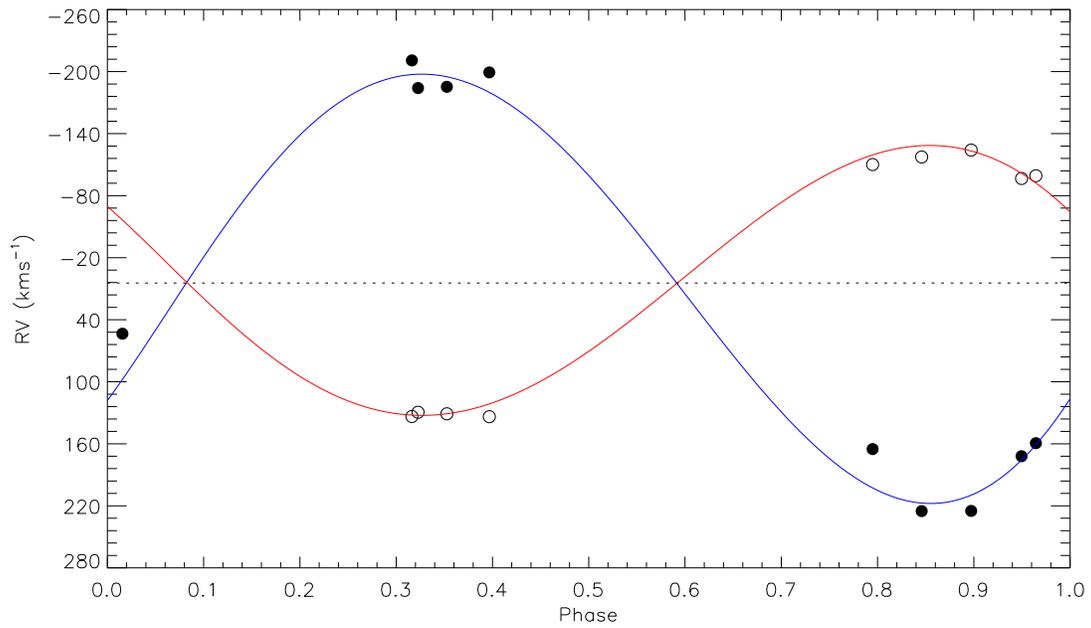}
\vspace{0.1 cm}
\caption{Radial velocity curve of HH\,Boo. Open circles represent the observations of the primary, while filled circles represent the secondary components. Solid curves are the theoretical radial velocity curves derived by the Spectroscopic Binary Solver software.}
\label{Fig.2}
\end{figure*}

\begin{figure*}
\hspace{2.1 cm}
\includegraphics[width=20cm]{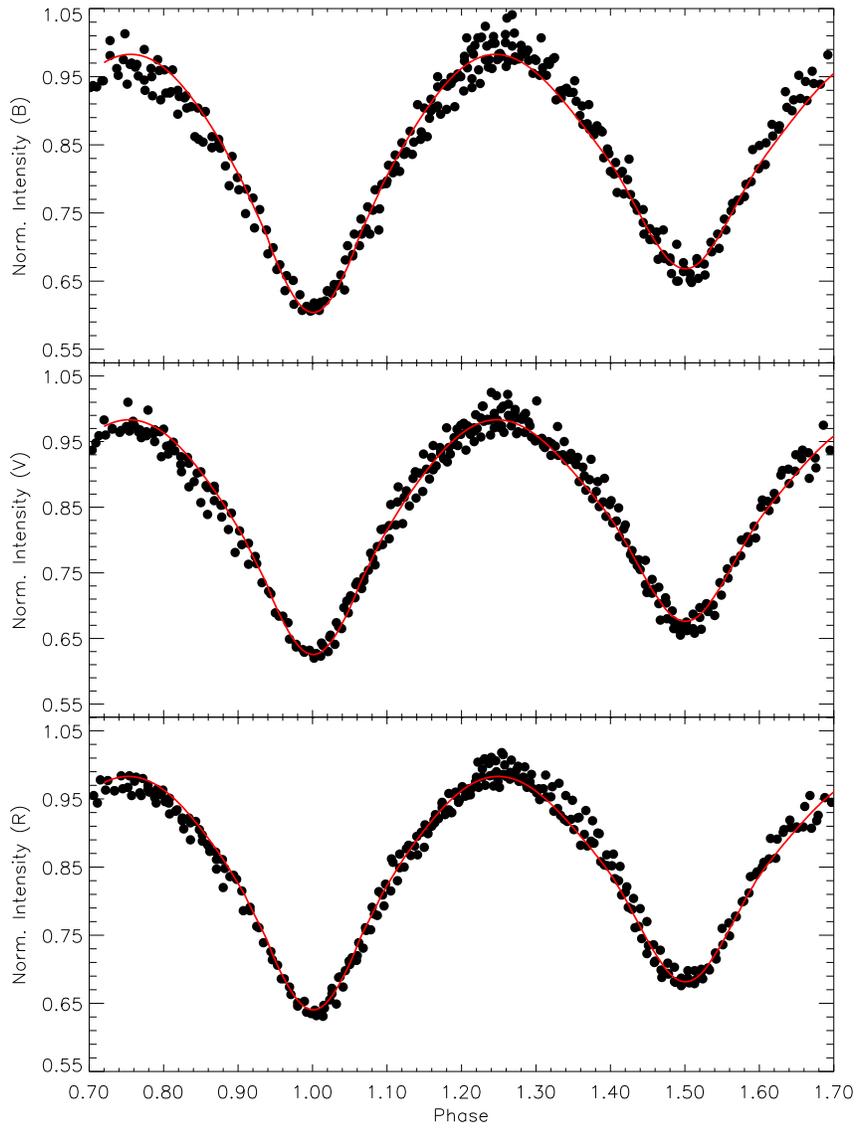}
\vspace{0.1 cm}
\caption{The BVR light curves of HH\,Boo and the synthetic solutions for the observations in each band.}
\label{Fig.3}
\end{figure*}

\begin{figure*}
\hspace{0.0 cm}
\includegraphics[width=15cm]{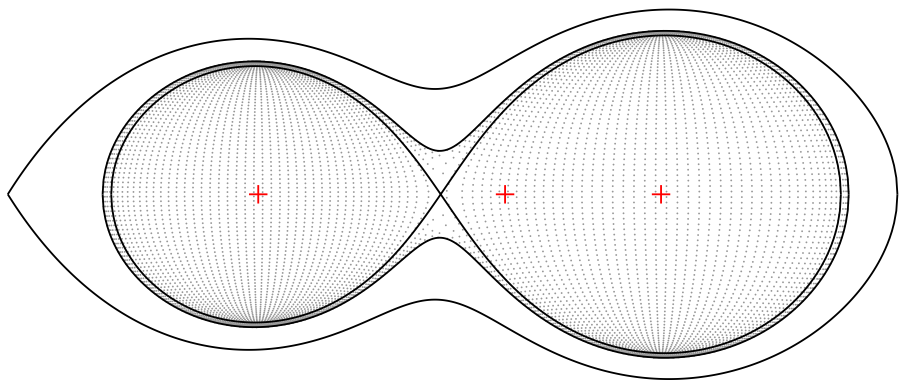}
\vspace{-8.0 cm}
\caption{The Roche geometry of HH\,Boo.}
\label{Fig.4}
\end{figure*}

\begin{figure*}
\hspace{2.5 cm}
\includegraphics[width=22cm]{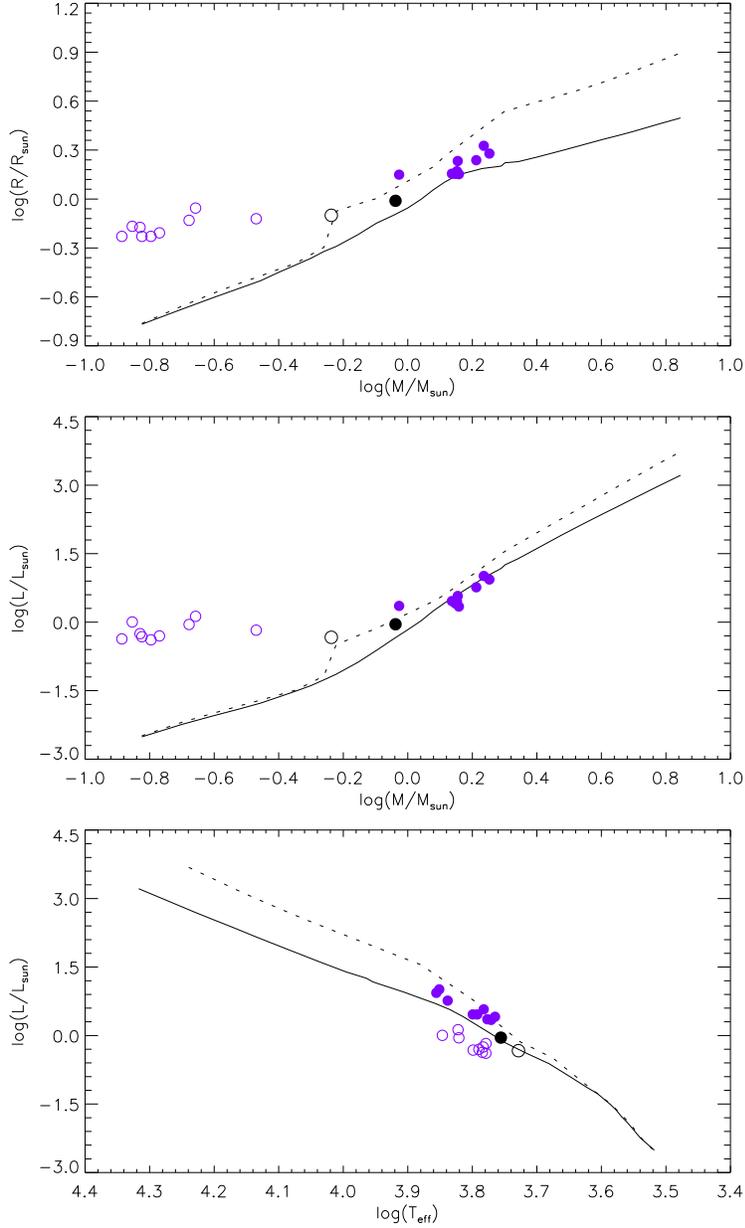}
\vspace{0.1 cm}
\caption{The places of the components of HH\,Boo in the planes of (upper panel) the mass-radius ($M-R$), (middle panel) mass-luminosity ($M-L$), and (bottom panel) luminosity-effective temperature ($T_{eff}-L$). In the panels, the continuous and dashed lines represent the ZAMS and TAMS theoretical models developed by \citet{Gir00}, respectively. The filled circles represent the primary components, while the open circles represent the secondary ones. The dark circles represent the HH\,Boo components, while the purple coloured circles represent the components of other contact binaries.}
\label{Fig.5}
\end{figure*}

\clearpage

\begin{table*}
\centering
\caption{Basic parameters for the observed stars.}
\vspace{0.3cm}
\begin{tabular}{lcccc}
\hline\hline
Star	&	Alpha (J2000)	&	Delta (J2000)	&	V	&	B-V	\\
	&	($^{h}$ $^{m}$ $^{s}$)	&	($^{\circ}$ $^{\prime}$ $^{\prime\prime}$)	&	(mag)	&	(mag)	\\
\hline									
HH\,Boo	&	14 21 44.06	&	46 41 59.40	&	11.272	&	0.654	\\
GSC\,03472-00043	&	14 21 46.59	&	46 43 49.40	&	12.906	&	0.595	\\
GSC\,03472-01201	&	14 21 37.30	&	46 46 18.90	&	13.701	&	0.662	\\
\hline
\end{tabular} 
\end{table*}

\begin{table*}
\centering
\caption{The minima times and ($O-C$) residuals (In the first column, the standard deviations of obtained minima times are given in the brackets near themselves).}
\vspace{0.3cm}
\begin{tabular}{lrcccr}
\hline\hline
O	&	E	&	$(O-C)_{II}$	&	Type	&	Method	&	REF	\\
\hline											
52749.5328	&	-9420.0	&	-0.00179	&	I	&	V	&	\citet{Pas06}	\\
52764.5096	&	-9373.0	&	-0.00230	&	I	&	V	&	\citet{Pas06}	\\
54148.6406	&	-5029.5	&	0.00192	&	II	&	Ir	&	\citet{Pas06}	\\
54513.5136	&	-3884.5	&	0.00209	&	II	&	Ir	&	\citet{Hub10}	\\
54599.3935	&	-3615.0	&	0.00146	&	I	&	CCD+R	&	\citet{Pas06}	\\
54912.3233	&	-2633.0	&	0.00100	&	I	&	Ir	&	\citet{Pas06}	\\
54912.4835	&	-2632.5	&	0.00187	&	II	&	Ir	&	\citet{Pas06}	\\
54937.3392	&	-2554.5	&	0.00160	&	II	&	Ir	&	\citet{Pas06}	\\
54958.6887	&	-2487.5	&	0.00046	&	II	&	V	&	\citet{Pas06}	\\
54958.8490	&	-2487.0	&	0.00143	&	I	&	V	&	\citet{Pas06}	\\
55644.9336	&	-334.0	&	-0.00237	&	I	&	V	&	\citet{Pas06}	\\
55680.7853	&	-221.5	&	-0.00062	&	II	&	V	&	\citet{Pas06}	\\
55716.4767(3)	&	-109.5	&	0.00013	&	II	&	R	&	This Study	\\
55716.4768(5)	&	-109.5	&	0.00023	&	II	&	B	&	This Study	\\
55716.4768(2)	&	-109.5	&	0.00025	&	II	&	V	&	This Study	\\
55743.4026(3)	&	-25.0	&	-0.00128	&	I	&	R	&	This Study	\\
55743.4032(4)	&	-25.0	&	-0.00060	&	I	&	V	&	This Study	\\
55743.4038(2)	&	-25.0	&	-0.00008	&	I	&	B	&	This Study	\\
55747.3863(5)	&	-12.5	&	-0.00083	&	II	&	R	&	This Study	\\
55747.3871(3)	&	-12.5	&	-0.00003	&	II	&	V	&	This Study	\\
55747.3874(4)	&	-12.5	&	0.00022	&	II	&	B	&	This Study	\\
55751.3696(3)	&	0.0	&	-0.00087	&	I	&	V	&	This Study	\\
55751.3700(3)	&	0.0	&	-0.00051	&	I	&	R	&	This Study	\\
55751.3700(6)	&	0.0	&	-0.00050	&	I	&	B	&	This Study	\\
55751.3701(2)	&	0.0	&	-0.00037	&	I	&	R	&	This Study	\\
55751.3702(3)	&	0.0	&	-0.00027	&	I	&	V	&	This Study	\\
55751.3702(4)	&	0.0	&	-0.00026	&	I	&	B	&	This Study	\\
\hline
\end{tabular} 
\end{table*}

\begin{table*}
\centering
\caption{The results of the analysis of Radial Velocity curve.}
\vspace{0.3cm}
\begin{tabular}{lrr}
\hline\hline
Parameter	&	Value	&	Error	\\
\hline					
Long. of Periastron$_{1}$ ($^\circ$)	&	238.669	&	$^{+49.884}_{-72.41}$	\\
Long. of Periastron$_{2}$ ($^\circ$)	&	58.669	&	$^{+49.884}_{-72.41}$	\\
Eccentricity ($e$)	&	0.048	&	$^{+0.043}_{-0.042}$	\\
Semi-Amplitude$_{1}$ ($kms^{-1}$)	&	129.647	&	$^{+6.4243}_{-6.2471}$	\\
Semi-Amplitude$_{2}$ ($kms^{-1}$)	&	205.078	&	$^{+7.0675}_{-6.7157}$	\\
Systemic Velocity ($kms^{-1}$)	&	2.9209	&	$^{+4.0328}_{-4.0429}$	\\
$a_{1}sin i$ ($km$)	&	$5.67\times10^{5}$	&	$\pm3.68\times10^{4}$	\\
$a_{2}sin i$ ($km$)	&	$8.98\times10^{5}$	&	$\pm4.86\times10^{4}$	\\
$m_{1}sin^{3} i$ ($M_{\odot}$)	&	$7.56\times10^{-1}$	&	$\pm9.99\times10^{-2}$	\\
$m_{2}sin^{3} i$ ($M_{\odot}$)	&	$4.78\times10^{-1}$	&	$\pm6.54\times10^{-2}$	\\
\hline
\end{tabular} 
\end{table*}

\begin{table*}
\centering
\caption{The parameters of components obtained from the light curve analysis.}
\vspace{0.3cm}
\begin{tabular}{lr}
\hline\hline
Parameter	&	Value	\\
\hline			
$i$ ($^\circ$)	&	69.71$\pm$0.16	\\
$T_{1}$ (K)	&	5699	\\
$T_{2}$ (K)	&	5352$\pm$15	\\
$\Omega_{1}$	&	3.0492	\\
$\Omega_{2}$	&	3.0402$\pm$0.0051	\\
$L_{1}/L_{T}$ (B)	&	0.691$\pm$0.046	\\
$L_{1}/L_{T}$ (V)	&	0.669$\pm$0.037	\\
$L_{1}/L_{T}$ (R)	&	0.657$\pm$0.031	\\
$g_{1}$, $g_{2}$	&	0.32, 0.32	\\
$A_{1}$, $A_{2}$	&	0.50, 0.50	\\
$x_{1,bol}$, $x_{2,bol}$	&	0.619, 0.619	\\
$x_{1,B}$, $x_{2,B}$	&	0.765, 0.765	\\
$x_{1,V}$, $x_{2,V}$	&	0.732, 0.732	\\
$x_{1,R}$, $x_{2,R}$	&	0.655, 0.655	\\
$<r_{1}>$	&	0.435$\pm$0.001	\\
$<r_{2}>$	&	0.354$\pm$0.001	\\
\hline
\end{tabular} 
\end{table*}

\end{document}